\def\ie{{\it i.e.}}

\def\o{{\cal O}}
\def\pk{P_k}
\def\Det{{\rm Det}}
\def\Tr{{\rm Tr}}
\def\det{{\rm det}}
\def\tr{{\rm tr}}
\def\dphi{{d_\phi}}
\def\dlambda{{d_\lambda}}
\def\lap{{\nabla^2}}
\font\titlefont=cmbx10 scaled\magstep1
\null
\rightline{SISSA 117/97/EP}
\rightline{\tt hep-th 9707239}
\null
\vskip 3cm
\centerline{\titlefont THE RUNNING GRAVITATIONAL COUPLINGS}
\smallskip
\vskip 3cm
\centerline{ \bf{ Djamel Dou$^{a,c}$ and Roberto Percacci$^{a,b}$ }}
\smallskip
\centerline{$^a$ International School for Advanced Studies, via Beirut 4}
\centerline{ 34014 Trieste, Italy}
\centerline{$^b$ Istituto Nazionale di Fisica Nucleare, 
Sezione di Trieste}

\smallskip
\centerline{$^c$  Institute of Theoretical Physics, Constantine University}
\centerline{Constantine, Algeria}

\vskip 1.8cm
\centerline{\bf Abstract}
\midinsert
\narrower\narrower\noindent
We compute the running of the cosmological constant and Newton's constant
taking into account the effect of quantum fields with any spin between 0 
and 2. We find that Newton's constant does not vary appreciably but
the cosmological constant can change by many orders of magnitude when one
goes from cosmological scales to typical elementary particle scales.
In the extreme infrared, zero modes drive the cosmological constant to zero.
\endinsert
\bigskip
\vfil
\eject
\leftline{\bf 1. Introduction}
\smallskip
\noindent
It is widely accepted that General Relativity (GR) can be regarded as a
low energy effective theory and can be treated as a quantum field theory
with an ultraviolet cutoff of the order of Planck's energy
(see [1] and references therein).
When energies of this magnitude are approached, corrections to
Einstein's theory with terms quadratic or of higher order in the curvature
will become relevant. Eventually, ``new physics'' is expected to set in.
In the low energy limit the terms with the lowest number of derivatives of 
the fields dominate; assuming the effective action to be described by a
local functional of the metric, the cosmological term will be the dominant one.
Here one faces the problem of explaining why this term should be so small
in natural units. For a general review of the problem see [2]. 
Several possible mechanisms involving quantum effects have been proposed
[3,4,5].
One expects that the cosmological constant problem can be solved within
the context of the effective theory, without reference to ultrahigh
energy physics. 

In this paper we study the running of Newton's constant $G=1/16\pi\kappa$
and of the cosmological constant $\Lambda$.
We shall consider ``matter'' fields coupled minimally to a
background metric; as usual in the background field method, the graviton 
(spin 2) field is treated like another matter field. All interactions between
matter fields will be neglected, except for their interaction with the metric.
The only self-interacting field will be the
graviton, which in accordance with the discussion above is supposed to
have an action consisting of a cosmological and Einstein term.

We use the Wilsonian approach to the renormalization group,
in a formulation given by Wetterich [6] and later
applied in a gravitational context by Reuter [7]. 
In this approach one writes an exact renormalization group equation
for a coarse-grained effective action, depending
on some external energy scale $k$. This exact equation is then
approximated by postulating a specific form for the action, containing 
only a few terms. 
In this way we obtain RG equations for $\kappa$ and $\Lambda$, 
containing contributions from all possible matter fields with spin
up to two. These solutions can be solved in the high-energy (low curvature)
and low energy (high curvature) approximations.
In the high energy regime the behaviour of the couplings follows power laws
dictated essentially by dimensional arguments. In the low energy regime 
the running of the couplings essentially stops, except for a residual
effect of the zero modes, which drives the ratio $\Lambda/\kappa$ to zero.

The plan of the paper is as follows: in section 2 we write down the
exact equation, in section 3 we make an Ansatz for the effective action
and derive RG equations for the cosmological constant and Newton's constant;
in sections 4 and 5 we study the high- and low-energy limits respectively;
in section 6 we extract gauge-independent information from our calculations.
In section 7 we discuss the significance and the limitations of our 
results, and their relation to other works.
\medskip

\leftline{\bf 2. The exact RG}
\smallskip
\noindent

In a scalar field theory with quantum fields $\phi$ propagating
on a Euclidean background metric $g_{\mu\nu}$ we define
a modified generating functional for connected Green functions, $W_k$,
by adding to the classical action a term $\Delta S_k$, depending on an external
momentum scale $k$:
$$
e^{-W_k(j)}=\int (d\phi)e^{-S(\phi)-\int j\phi-\Delta S_k(\phi)}\ .
\eqno(2.1)
$$
The new term is constructed so as to suppress the propagation of modes with
momenta smaller than $k$. We will refer to it as an infrared cutoff.
It can be taken purely quadratic in the fields, with the general form:
$$
\Delta S_k(\phi)={1\over 2}\int d^4x \sqrt g 
\phi\, R_k\left(-\lap\right)\phi \ ,
\eqno(2.2)
$$
where $R_k=k^2\,F\left({-\lap\over k^2}\right)$ is a suitable 
function of the Laplacian.  
For example one may consider exponential cutoff functions of
the form
$$
R_k(z)={z f_k^2(z)\over 1-f_k^2(z)}\ ,\eqno(2.3)
$$
where
$$
f_k(z)=\exp\left(-a\left({z\over k^2}\right)^b\right)\ .\eqno(2.4)
$$
We allow $b$ to have values between 1 (gaussian cutoff) and infinity
(sharp cutoff). In general the results will depend on the form of
the cutoff function, i.e. on the parameters $a$ and $b$.

In flat space, the effect of the addition of the term $\Delta S_k$
is to replace in the propagators the momentum squared by the modified
inverse propagator:
$$
q^2+R_k(q^2)={q^2\over 1-f_k^2(q^2)}=P_k(q^2)\ .\eqno(2.5)
$$
The effective average action is defined as the Legendre transform of 
$W_k$, minus the cutoff:
$$
\Gamma_k(\phi)=W_k(j)-\int d^4x j\phi-\Delta S_k(\phi)\ .\eqno(2.6)
$$
Heuristically, one can interpret $\Gamma_k$ as a kind of coarse-grained
action. The scale $k$ can be understood physically 
as the resolution with which the system is observed by some apparatus.
Modes with momenta larger than $k$ cannot be directly observed and their
effect is averaged out by the functional integral.
The fact that certain results depend on the shape of the cutoff function
just means that observing the system in different ways will show
different pictures. Other results may turn out to be independent of the
shape of the cutoff function; they are then said to be universal.

One can easily prove that the average effective action satisfies the following
exact renormalization group equation [6]:
$$
\partial_t\Gamma_k={1\over 2} \Tr
\left({\delta^2 \Gamma_k\over\delta\phi\delta\phi}+R_k\right)^{-1}
\partial_t R_k\ .\eqno(2.7)
$$
This equation describes the flow of $\Gamma_k$ as the external parameter
$k$ is varied.
It is important to note that due to the form of the
cutoff term $\Delta S_k$, the trace is actually finite, so there is no need
for any regularization or renormalization to make sense of it.

The exact equation has much the same form as an improved
one-loop renormalization group equation. In fact, at one loop
the effective action would be
$$
\Gamma^{(1)}={1\over 2}\ln\det\o\ \ ;\qquad\qquad
\o={\delta^2 S\over\delta\phi\delta\phi}\eqno(2.8)
$$
and therefore if we define a modified one-loop effective action by
$$
\Gamma_k^{(1)}={1\over 2}\ln\det{\o_k}\ \ ; \qquad\qquad
\o_k={\delta^2(S+\Delta S_k)\over\delta\phi\delta\phi}=\o+R_k
\eqno(2.9)
$$
we get
$$
\partial_t\Gamma_k^{(1)}={1\over 2}(\det\o_k)^{-1}\partial_t\det\o_k
={1\over 2}\Tr\o_k^{-1}\partial_t\o_k\ ,\eqno(2.10)
$$
whence
$$
\partial_t\Gamma_k^{(1)}
={1\over 2} \Tr\left({\delta^2 S\over\delta\phi\delta\phi}+R_k\right)^{-1}
\partial_t R_k\ .\eqno(2.11)
$$
This is the same as (2.7), except for the replacement of the classical
action $S$ by $\Gamma_k^{(1)}$ at $\o^{-1}$: this is the ``renormalization 
group improvement''.

The exact equations for fields with higher spin can be obtained following
similar lines. The case of spin 1/2 fields was discussed in [8]
for a flat background. In a curved background the Dirac operator is
$\gamma^\mu\nabla_\mu$, where $\nabla_\mu\psi=\partial_\mu\psi
+\omega_\mu^{ab}T_{ab}\psi$, with the spin connection given by
$\omega_\mu^{ab}=e^a_\lambda \Gamma^\lambda_{\mu\nu}e^{b\nu}
+e^a_\lambda\partial_\mu e^{b\lambda}$ and
$T_{ab}={1\over8}[\gamma_a,\gamma_b]$. 
The square of the Dirac operator is $-\lap+{1\over4}R$. 
For Dirac spinors in a gravitational background,
writing $\Delta S_k=\int d^4x\,\bar\psi R_k^{(D)}(\gamma^\mu\nabla_\mu)\psi$ 
one obtains an exact RG equation of the form: 
$$
\partial_t\Gamma_k=-\Tr
\left({\delta^2 \Gamma_k\over\delta\bar\psi\delta\psi}+R_k^{(D)}\right)^{-1}
\partial_t R_k^{(D)}\ ,\eqno(2.12)
$$
with the trace now running also on the Dirac indices.
The same equation holds also for Weyl spinors, with the trace running over the
appropriate subspace.
In the case of Majorana spinors the RG equation has again the same form 
except for an overall factor 1/2 on the r.h.s.
For massless spin 1 fields there is the extra complication of
gauge invariance. This has been discussed in [9].
One can obtain an exact RG equation by interpreting $S$ as the complete
action, including gauge-fixing and ghost terms. 
For the photon field we will use a gauge fixing term of the form
$$
S_{\rm GF}(A,g)=
{1\over2\lambda} \int d^4x\,\sqrt{g}\ (\nabla^\mu A_\mu)^2\ ,
\eqno(2.13)
$$
with the corresponding ghost term
$$
S_{\rm ghost}(c,\bar c,g)=\int d^4x\ \sqrt{g}\,\bar c(-\lap)c\ .
\eqno(2.14)
$$
The resulting RG equation is
$$
\partial_t\Gamma_k={1\over 2} \Tr
\left({\delta^2 \Gamma_k\over\delta A\delta A}+R_k\right)^{-1}
\partial_t R_k
-\left({\delta^2 \Gamma_k\over\delta\bar c\delta c}+R_k^{\rm (gh)}\right)^{-1}
\partial_t R_k^{\rm (gh)}\ ,\eqno(2.15)
$$
In the case of a massive (Proca) vector field one has to drop the ghost term.

For nonchiral spin $3/2$ (Rarita-Schwinger) fields, the RG equation takes the 
same form as for Dirac spinors except
that there will be spacetime indices as well as Dirac indices.
We write it formally:
$$
\partial_t\Gamma_k=-\Tr
\left({\delta^2 \Gamma_k\over\delta\bar\Psi\delta\Psi}+R_k^{(RS)}\right)^{-1}
\partial_t R_k^{(RS)}\ ,\eqno(2.16)
$$

Finally, the RG equation for a spin 2 field has been discussed in [7].
For diffeomorphism invariance we choose a background gauge fixing with
background metric $g_{\mu\nu}$. Denoting $\gamma$ the dynamical metric,
$$
S_{\rm GF}(\gamma,g)=
{1\over2\alpha}Z_k \int d^4x\,\sqrt{g}g^{\mu\nu} F_\mu F_\nu\ .
\eqno(2.17)
$$
where 
$F_\nu=\sqrt{\bar\kappa}\left(\delta^\sigma_\nu\nabla^\rho 
-{1\over2}g^{\rho\sigma}\nabla_\nu\right)
(\gamma_{\rho\sigma}-g_{\rho\sigma})$ and $\nabla$ denotes the
covariant derivative with respect to the background metric.
The constant $Z_k$ can be written as $Z_k=\kappa_k/\bar\kappa$, where
$\kappa_k$ is the scale-dependent Newton constant and
$\bar\kappa$ is the running Newton constant evaluated at some 
arbitrary reference energy scale $\bar k$ (see eqns.(3.3) and (3.19) below).  
The corresponding ghost term is 
$$
S_{\rm ghost}(v,\bar v,g)=\int d^4x\ \sqrt{g}\ \bar v_\mu
\left(-g^{\mu\nu}\lap-R^{\mu\nu}\right) v_\nu
\eqno(2.18)
$$
The exact equation has the form:
$$
\partial_t\Gamma_k={1\over 2} \Tr
\left({\delta^2 \Gamma_k\over\delta h\delta h}+R_k\right)^{-1}
\partial_t R_k
-\left({\delta^2 \Gamma_k\over\delta\bar v\delta v}+R_k^{\rm (gh)}\right)^{-1}
\partial_t R_k^{(\rm gh)}\ ,\eqno(2.19)
$$
where $\gamma_{\mu\nu}=g_{\mu\nu}+\sqrt{\bar\kappa}h_{\mu\nu}$.

Finally, we can consider a theory where all the fields mentioned so
far are present at the same time. Assuming that the (quadratic)
cutoff terms do not mix fields with different spins, the RG equation for the
coupled system has the form
$$
\partial_t\Gamma_k=
\partial_t\Gamma_k|_{\rm scalar}
+\partial_t\Gamma_k|_{\rm spinor}
+\partial_t\Gamma_k|_{\rm vector}
+\partial_t\Gamma_k|_{\rm RS}
+\partial_t\Gamma_k|_{\rm gravity}
\eqno(2.20)
$$
where the terms on the r.h.s. are given by equations 
(2.11), (2.12), (2.15), (2.16), (2.19).

\goodbreak
\medskip
\leftline{\bf 3. The truncated gravitational RG}
\smallskip
\noindent
The exact equation (2.20) describes a flow in the infinite
dimensional space of all possible action functionals for the theory.
It is therefore equivalent to infinitely many equations in
infinitely many unknowns (the couplings of all possible terms in the
Lagrangian). It is an impossible task to solve this equation.

The simplest thing to do is to assume that the action has a certain form,
parametrized by a finite number of coupling constants, plug it in the
equation and solve the resulting finite system of PDE's.
The virtue of this approach is that the truncated, finite system
still contains genuine nonperturbative information. The flip side
is that there is no control on the approximation.

We separate the gravitational action, whose parameters we want to
study, from the matter action, which has a fixed form and is not subject to
RG flow:
$$
\Gamma_k=\Gamma_k^{\rm matter}+\Gamma_k^{\rm gravity}\ .
\eqno(3.1)
$$
For the matter sector we consider ``free'' matter fields,
i.e. matter fields that are coupled minimally to the metric but otherwise have
no other interactions. We will consider $n_S$ scalar fields $\phi$, 
$n_D$ Dirac fields $\psi$, $n_M$ Maxwell fields
$A_\mu$, $n_P$ (Proca) massive vector fields and $n_{RS}$ Rarita-Schwinger
fields. Here the spinor fields are supposed massive and therefore 
non-chiral. The results for chiral spinors are obtained replacing $n_D$
by $n_W/2$ ($n_W$ being the number of Weyl spinors) and $n_{RS}$
by $n_{CRS}/2$ ($n_{CRS}$ being the number of chiral Rarita-Schwinger fields).

The action for these fields is taken to be:  
$$
\eqalign{
\Gamma_k^{\rm matter}=\sum\int d^4x\sqrt \gamma \Biggl[
&
{1\over2}\phi\left(-\lap+m_S^2\right)\phi
+{\bar\psi}\left(\gamma^\mu\nabla_\mu+m_D\right)\psi
+{1\over 4}F^{\mu\nu}F_{\mu\nu}
+{1\over2\lambda}\left(\nabla_\nu A^\nu\right)^2 
+\bar c(-\lap) c
\cr
&+{1\over 4}G^{\mu\nu}G_{\mu\nu}
+{1\over2}m_P^2 B_\mu B^\mu 
+{1\over2}\epsilon^{\mu\nu\rho\sigma}
{\bar\Psi}_\mu\gamma_5\gamma_\nu \nabla_\rho\Psi_\sigma
+m_{RS}{\bar\Psi}_\mu \sigma_{\mu\nu} \Psi_\nu
\Biggr] \ , \cr}
\eqno(3.2)
$$
the sum being over all particle species. The tensors $F_{\mu\nu}$
and $G_{\mu\nu}$ are the field strengths for $A_\mu$ and $B_\mu$
respectively and all covariant derivatives are calculated with respect to the
metric $\gamma$. For the gravitational action we take
$$
\Gamma_k^{\rm gravity}(\gamma,g)=\kappa_k\int
d^4x\sqrt{\gamma}\left(2\Lambda_k-R(\gamma)\right) 
+S_{\rm GF}(\gamma,g)+S_{\rm ghost}(v,\bar v,g)\ ,
\eqno(3.3) $$
where $\kappa_k=1/16\pi G_k$, $\Lambda_k$ is the running cosmological 
constant, $G_k$ is the running Newton constant and the gauge fixing and
ghost terms are as in (2.17) and (2.18). The background metric appears
only in the gauge fixing term.

As already remarked, the contributions of different fields are all
decoupled. Given the explicit form of the action (3.2) and (3.3) we now
have to
evaluate the second derivatives that appear on the r.h.s. of equations
(2.11), (2.12), (2.15), (2.16) and (2.19), specify the cutoff functions
$R_k$ and evaluate the traces.

Since we want to isolate the coefficients of terms containing zero and one 
power of the curvature, it is enough to evaluate the r.h.s. for a background
geometry of constant curvature, namely one satisfying:
$$
R_{\mu\nu\rho\sigma}={1\over12}R
(g_{\mu\rho}g_{\nu\sigma}-g_{\mu\sigma}g_{\nu\rho})\ ;\qquad
R_{\mu\nu}={1\over4}R g_{\mu\nu}\ .
\eqno(3.4)
$$
The scalar curvature $R$ can be either positive or negative, corresponding
to de Euclidean Sitter or anti-de Sitter space.
We also assume that all matter fields have vanishing vacuum expectation
value, and therefore vanishing energy-momentum tensor.
With this Ansatz, the classical field equation coming from (3.1-2-3)
just says that $R=4\Lambda_k$. However, we shall stay off-shell as long
as possible and therefore will not assume this equation for the
background for the time being.

For $n_S$ scalar fields we have simply:
$$
\partial_t\Gamma_k|_{\rm scalar}=
{n_S\over2}\Tr_{(0)}{\partial_t\pk\over\pk+m_S^2}\ .
\eqno(3.5)
$$
The suffix $(0)$ is a reminder that the trace is on a field of spin zero.

For Dirac spinor fields we choose the cutoff term in the form 
$$
\Delta S_k=\int d^4x\, \det e\, \bar\psi R_k^{(D)}(\gamma^\mu\nabla_\mu)\psi\ ,
\eqno(3.6)
$$
where $e^a{}_\mu$ is the vierbein for the metric $g$ 
and the function $R_k^{(D)}$ is
$$
R_k^{(D)}(\gamma^\mu\nabla_\mu)=\left({\sqrt{P_k(-\lap)+{R\over4}}\over
\sqrt{-\lap+{R\over4}}}-1\right) \ \gamma^\mu\nabla_\mu\ .
\eqno(3.7)
$$
This results in the replacement of the Dirac operator 
$\gamma^\mu\nabla_\mu$ by 
${\sqrt{P_k+{R\over4}}\over\sqrt{-\lap+{R\over4}}}\gamma^\mu\nabla_\mu$,
whose square is $P_k+{R\over4}$. 
(In flat space, $q_\mu$ is replaced by $\sqrt{\pk(q^2)}\hat q_\mu$,
where $\hat q_\mu=q_\mu/|q|$.) Note that there are no ordering ambiguities
in (3.7), as $\gamma^\mu\nabla_\mu$ commutes with $-\lap$ on de Sitter space.
\footnote{$^*$}{Alternatively we could have chosen 
$\sqrt{P_k\left(-\lap+R/4\right)}$ in the numerator of (3.7).
The square of the modified Dirac operator would then be 
$\pk\left(-\lap+R/4\right)$. This operator commutes with
$\gamma^\mu\nabla_\mu$ for any background field.}
For $n_D$ Dirac spinors we then have
$$
\partial_t\Gamma_k|_{\rm spinor}=
-{n_D\over2}\Tr_{(1/2)}{\partial_t\pk\over\pk+{R\over4}+m_D^2}\ ,
\eqno(3.8)
$$
the trace being over four-component Dirac spinors. For Weyl and Majorana 
spinors the remarks following eq.(2.12) apply.

For fields of higher spin, carrying reducible representations of $SO(4)$,
it turns out to be convenient to decompose further each type
of field into its irreducible components.
Actually, on Euclidean de Sitter space, one can further decompose
in representations of $SO(5)$ [10,11].
For a vector field $A_\mu$ we use the decomposition
$A_\mu=A_\mu^T+\nabla_\mu\phi$, where $\nabla^\mu A_\mu^T=0$.
In the functional integral this change of variables gives rise to a Jacobian
determinant, which is $J=\sqrt{\Det\left(-\lap\right)}$ [11].
For reasons of dimensional homogeneity it is convenient to define
$\hat\phi=\sqrt{(-\lap)}\phi$. The Jacobian for this further change of
variables cancels the preceding one. 
\footnote{$^\dagger$}
{We observe that if we were to
work with $\phi$ instead, one should not introduce a cutoff term
in the Jacobian determinant. This alternative procedure would lead to the same
results as the method followed here.}
Taking into account also the gauge fixing term (2.13) we have
for massless fields 
$$
\Gamma_k^{(2)}={n_M\over2}\int d^4x\sqrt g
\left[A_\mu^T(-\lap+{R\over4})g^{\mu\nu}A_\nu^T
+{1\over\lambda}\hat\phi(-\lap)\hat\phi\right]
\eqno(3.9)
$$
whereas for massive fields
$$
\Gamma_k^{(2)}={n_P\over2}\int d^4x \sqrt g
\left[A_\mu^T(-\lap+{R\over4}+m_P^2)g^{\mu\nu}A_\nu^T
-m_p^2\hat\phi^2\right]\ .
\eqno(3.10)
$$
In evaluating the trace that appears in Eq.(2.15) one has to pay special
attention to zero modes. In fact, the operator $-\lap$ has a constant
zero mode when acting on scalars which does not actually correspond to any 
eigenvalue of $-\lap$ on $A_\mu$. This zero mode has therefore to be
removed from the trace. Similarly, the ghost zero mode does not correspond
to any gauge transformation of $A_\mu$ and should also not be counted [4].
The cutoff is chosen as for scalar fields by requiring the replacement of 
$-\lap$ with $\pk(-\lap)$. The ghost contribution partly cancels
the longitudinal photon contribution, and we find
$$
\partial_t\Gamma_k|_{\rm vector}=
{1\over2}n_M\left[\Tr_{(1)}{\partial_t\pk\over \pk+{R\over4}}
-\Tr_{(0)}^\prime{\partial_t\pk\over\pk}\right]
+{1\over2}n_P\Tr_{(1)}{\partial_t\pk\over \pk+{R\over4}+m_P^2}\ ,
\eqno(3.11)
$$
where $\Tr^\prime$ is a sum over all nonzero modes.

For the Rarita-Schwinger field one has the decomposition:
$$
\eqalign{
\Psi_\mu=&\Phi_\mu+{1\over4}\gamma_\mu \psi\ ;
\qquad \gamma^\mu\Phi_\mu=0\cr
\Phi_\mu=&\Phi_\mu^T+\left(\nabla_\mu-{1\over4}\gamma_\mu\gamma^\nu\nabla_\nu\right)
\zeta\ ;\qquad \nabla^\mu\Phi_\mu^T=0
\cr}\eqno(3.12)
$$
with the corresponding functional Jacobian
$\sqrt{\Det\left(-\lap-{1\over12}R\right)}$ [11].
Again it is convenient to define $\hat\zeta=\sqrt{-\lap-{1\over12}R}\zeta$,
so as to deal with fields all having the same (canonical) dimension
and to cancel the functional Jacobian in the path integral.
The linearized action is
$$
\eqalign{
\Gamma_k^{(2)}\!=&{n_{RS}\over2}\!\int\! d^4x \sqrt g
\Biggl\{
\Phi^{\mu T}(\gamma^\mu\nabla_\mu-m_{RS})\Phi_\nu^T\cr
&
+{3\over16}\left[\bar\zeta\left(\gamma^\mu\nabla_\nu+2m_{RS}\right)\zeta
-2\bar\zeta\sqrt{-\lap-{R\over12}}\psi
-\bar\psi\left(\gamma^\mu\nabla_\mu-2m_{RS}\right)\psi\right]
\Biggr\}\ .\cr}
\eqno(3.13)
$$
The square of the operator acting on $\Phi_\mu^T$ is
$-\lap+{R\over3}+m_{RS}^2$.
The cutoff is chosen as follows:
$$
\eqalign{
\Delta S_k\!=&{1\over2}n_{RS}\!\int\! d^4x \sqrt g
\Biggl\{
\Phi^{\mu T} R_k^{(RS)} \Phi_\nu^T
\cr
&\qquad+{3\over16}
\left[
\bar\zeta R_k^{(D)}\zeta
-2\bar\zeta\left(\sqrt{\pk-{R\over12}}-\sqrt{-\lap-{R\over12}}\right)\psi
-\bar\psi R_k^{(D)}\psi
\right]\Biggr\}\ ,\cr}
\eqno(3.14)
$$
where
$$
R_k^{(RS)}(\gamma^\mu\nabla_\mu)=\left({\sqrt{P_k(-\lap)+{R\over3}}\over
\sqrt{-\lap+{R\over3}}}-1\right) \gamma^\mu\nabla_\mu\ .
\eqno(3.15)
$$
Then $\Gamma_k^{(2)}+\Delta S_k$ is equal to (3.13) except for the
replacement of $-\lap$ by $\pk(-\lap)$. The r.h.s. of (2.16) becomes
$$
\partial_t\Gamma_k|_{RS}=
-{n_{RS}\over2}\Tr_{(3/2)}{\partial_t\pk\over\pk+{R\over3}+m_{RS}^2}\ .
\eqno(3.16)
$$

Finally, we come to the gravitons.
As usual we define the graviton field $h_{\mu\nu}$ by
$\gamma_{\mu\nu}=g_{\mu\nu}+ \sqrt{\bar\kappa}h_{\mu\nu}$.
The quantum fluctuation $h_{\mu\nu}$ is 
decomposed into irreducible parts under $SO(5)$:
$$
h_{\mu\nu}=h^T_{\mu\nu}+\nabla_\mu\xi_\nu+\nabla_\nu\xi_\mu
+\nabla_\nu\nabla_\nu\sigma-{1\over4}g_{\mu\nu}\nabla_\lambda\nabla^\lambda
\sigma+{1\over4} g_{\mu\nu}h\ ,
\eqno(3.17)
$$
where $h=g^{\mu\nu}h_{\mu\nu}$, $\xi_\mu$ satisfies 
$\nabla^\lambda\xi_\lambda=0$ and
$h^T_{\mu\nu}$ satisfies $g^{\mu\nu}h^T_{\mu\nu}=0$ 
and $\nabla^\lambda h^T_{\lambda\nu}=0$.

We use instead of $\xi$ and $\sigma$ the variables 
$\hat\xi_\mu=\sqrt{-\lap-{1\over4} R}\,\xi_\mu$
and $\hat\sigma=\sqrt{-\lap}\sqrt{-\lap-{1\over 3}R}\,\sigma$.
As in the previous cases, the Jacobian for the functional change of variables
from $h_{\mu\nu}$ to $h^T_{\mu\nu}$, $\xi_\mu$, $\sigma$, $h$,
$$
J=\Det_1\left(-\lap-{1\over4}R\right)^{1/2}
\Det_0\left(-\lap-{1\over3}R\right)^{1/2}
\Det_0\left(-\lap\right)^{1/2}
\eqno(3.18)
$$
exactly cancels the Jacobian for the functional change of variables
from $h^T_{\mu\nu}$, $\xi_\mu$, $\sigma$, $h$, 
to $h^T_{\mu\nu}$, $\hat\xi_\mu$, $\hat\sigma$, $h$.
Thus there is no functional Jacobian to be taken into account when
going from $h_{\mu\nu}$ to $h^T_{\mu\nu}$, $\hat\xi_\mu$, $\hat\sigma$, $h$.

The quadratic approximation to the effective action in the neighbourhood
of $g$, taking into account also the gauge fixing terms, has the form:
$$
\eqalign{
\Gamma_k^{(2)}={1\over2}Z_k\int d^4 x \sqrt{g}&\Biggl\{
{1\over2}
h^T_{\mu\nu}\left(-\lap+{2\over3}R-2\Lambda_k\right)h^{T\mu\nu} 
+{1\over\alpha}\hat\xi_\mu
\left(-\lap-{1-2\alpha\over4}R-2\alpha\Lambda_k\right)\hat\xi^\mu \cr
+&{3\over16}{3-\alpha\over\alpha}
\Biggl[
\hat\sigma\left(-\lap-{\alpha-1\over\alpha-3}R
+{4\alpha\over\alpha-3}\Lambda_k\right)\hat\sigma \cr
&\qquad\qquad +2{\alpha-1\over\alpha-3}\hat\sigma
\sqrt{-\lap}\sqrt{-\lap-{1\over3}R}\;h
-{3\alpha-1\over 3(3-\alpha)}
h\left(-\lap-{4\alpha\over3\alpha-1}\Lambda_k\right)h\Biggr]\Biggr\}\cr}.
\eqno(3.19)
$$
Note that $Z_k=\kappa_k/\bar\kappa$ 
plays the role of a wave function renormalization constant
for the graviton. The running of Newton's constant can therefore be
interpreted as due to an anomalous dimension of the graviton field.

The ghost and antighost fields can be decomposed as the vector field $A_\mu$;
the ghost term (2.18) becomes
$$
\Gamma_k^{(2)}={1\over2}\int d^4x\sqrt g
\left[\bar v_\mu^T(-\lap-{R\over4})g^{\mu\nu}v_\nu^T
+\hat\phi(-\lap-{R\over2})\hat\phi\right]\ .
\eqno(3.20)
$$
Note that we put the ghost wave function renormalization constant
equal to one.

We now have to decide the form of the cutoff terms $\Delta S_k(g)$.
As before we assume that all the states with definite spin 
and parity, including the ghosts, 
have their propagators modified by the replacement of
$-\lap$ with $\pk(-\lap)$. This is achieved by the following choice:
$$
\eqalign{
\Delta S_k={1\over2}Z_k\int d^4 x \sqrt{g}&\Biggl\{
{1\over2}
h^T_{\mu\nu}\left(\pk+\lap\right)h^{T\mu\nu}
+{1\over\alpha}\hat\xi_\mu\left(\pk+\lap\right)\hat\xi^\mu
+{3\over16}{3-\alpha\over\alpha}
\Biggl[
\hat\sigma\left(\pk+\lap\right)\hat\sigma \cr
&+2{\alpha-1\over\alpha-3}\hat\sigma
\left(\sqrt{\pk}\sqrt{\pk-{1\over3}R}-
\sqrt{-\lap}\sqrt{-\lap-{1\over3}R}\right)h
-{3\alpha-1\over 3(3-\alpha)}
h\left(\pk+\lap\right)h\Biggr]\Biggr\}\cr}
\eqno(3.21)
$$
so that $S_k+\Delta S_k$ is identical to (3.19) except for the substitution
of $-\lap$ by $\pk$ everywhere.

We now have to insert these formulae in the r.h.s. of (2.19).
We decompose all the fields into normal modes of the operator $-\lap$.
Things go slightly differently depending on the sign of the curvature.
For de Sitter space one has to take into account the fact that the first 
two normal modes 
of the field $\sigma$ do not contribute to $h_{\mu\nu}$ and therefore
have to be discarded (see also Appendix A). On the other hand even a constant
$h$ contributes to $h_{\mu\nu}$ and therefore all the modes of $h$ have to be
included. In this way, the first two modes of $h$
($i=0,1$ in the terminology of Table A.I) remain singled out in the trace [12].
The $t$-derivative acts on the explicit $k$-dependence in $\pk$ 
and also on the running parameter $Z_k$ (we shall see {\it a posteriori} that
this dependence is actually very weak and can be neglected).

Putting all together and taking the traces leads to the following:
$$
\eqalign{
\partial_t&\Gamma_k|_{\rm gravity}=
{1\over2}\Tr_{(2)}{\partial_t \pk\over \pk+{2\over3}R-2\Lambda_k}
+{1\over2}\Tr_{(1)}^\prime{\partial_t \pk\over 
\pk+{2\alpha-1\over4}R-2\alpha\Lambda_k} 
+{1\over2}\Tr_{(0)}^{\prime\prime}{\partial_t \pk\over
\pk+{\alpha-1\over2}R-2\alpha\Lambda_k} \cr &
+{1\over2}\Tr_{(0)}^{\prime\prime}{\partial_t \pk\over \pk-2\Lambda_k}
+{1\over2}\sum_{j=0,1}
{\partial_t\pk(\lambda^{(0)}_j)\over \pk(\lambda^{(0)}_j)
-{4\alpha\over 3\alpha-1}\Lambda_k}
-\Tr_{(1)}{\partial_t \pk\over \pk-{1\over4}R}
-\Tr_{(0)}^\prime{\partial_t \pk\over \pk-{1\over2}R} \cr
&+{\partial_t Z_k\over Z_k}
\Biggl\{
{1\over2}\Tr_{(2)}{\pk+\lap\over \pk+{2\over3}R-2\Lambda_k}
+{1\over2}\Tr_{(1)}^\prime{\pk+\lap\over
\pk+{2\alpha-1\over4}R-2\alpha\Lambda_k} 
+{1\over2}\sum_{j=0,1}
{\pk(\lambda^{(0)}_j)-\lambda^{(0)}_j\over \pk(\lambda^{(0)}_j)
-{4\alpha\over 3\alpha-1}\Lambda_k}
\cr
&-\!{1\over4\alpha}\Tr_{(0)}^{\prime\prime}{1\over 
(\pk-2\Lambda_k)\left(\pk+{\alpha-1\over2}R -2\alpha\Lambda_k\right)}
\Biggl[
\!\Bigl(\!(1-3\alpha)\bigl((3-\alpha)\pk\!+\!(\alpha-1)R/2\bigr)
\!+\!4\alpha(\alpha+1)\Lambda_k\!\Bigr)\!
(\pk+\lap) \cr
&\qquad\qquad\qquad-3(1-\alpha)^2\sqrt{\pk}\sqrt{\pk-{1\over3}R}
\left(\sqrt{\pk}\sqrt{\pk-{1\over3}R}-
\sqrt{-\lap}\sqrt{-\lap-{1\over3}R}\right)\Biggr]
\Biggr\}\cr}
\eqno(3.22)
$$
where a double prime means that the modes $i=0,1$ are omitted 
from the trace. In the case of anti-de Sitter space the spectrum
is continuous; the traces will be over all normal modes and the
terms involving isolated eigenmodes have to be dropped. 

The complete RG equation (2.20) is obtained by summing the r.h.s.
of eqns (3.5,8,11,16,22). 
There is obviously no exact solution even to this truncated equation.
In the next two sections we obtain approximate analytic results in the
two limiting cases: $k^2\gg |R|$ and $k^2\ll |R|$.
Before doing that, however, we can say something on
the general behaviour of the solutions, which can be guessed on the 
basis of dimensional considerations.

We will assume from now on that we are not too far off shell, 
\ie\  that $\Lambda_k$ is of the same order of magnitude as $R$. 
The most important contribution
to the traces comes from the modes with frequencies of order $k$.
This is because the functions $\partial_t P_k(z)$ and $P_k(z)-z$, decay
exponentially when $z\gg k^2$.
In the region $z\approx k^2$, $P_k(z)\approx k^2$, 
so if we assume that either $k^2\gg |R|$ or $k^2\ll |R|$,
in the denominators in (3.22) we can neglect $R$ and $\Lambda_k$
or $\pk$ respectively.

Similar considerations apply for the matter fields.
In addition, in the matter contributions there are various possible regimes,
depending on how the scale $k$ compares to the masses of the matter
fields. Generally speaking, if the mass of a matter field is 
greater than the scale $k$, that field decouples and gives a 
negligible contribution to the RG equation.
This can be seen for example in the case of a scalar field:
when $k^2\ll m_S^2$ we can neglect $P_k$ in the denominator of (3.5).
The remaining trace is proportional to $k^2$ for dimensional reasons, 
so the r.h.s. of (3.5) is of order $k^2/m_S^2\ll1$.
The situation is similar for the other fields.

As $k$ grows the RG equation passes through various mass thresholds, 
and every time
this happens the contribution of a new field has to be taken into account.
At a given energy scale $k$ therefore one need only consider the contributions
of fields with masses lower than $k$.
Thus for very low energies one needs to consider only the massless fields
will matter.

\medskip
\leftline{\bf 4. The small curvature (high energy) limit}
\smallskip
\noindent
In the case $k^2\gg R$ one is probing spacetime regions which are small
relative to the curvature radius of the background metric.
In such regions spacetime can be regarded as almost flat,
with small corrections. One can therefore reliably use the heat kernel
expansion to calculate the traces.
As noted in the end of the preceding section, the dominant contribution to
the traces comes from the region of momenta of order $k$, where $\pk$ is
itself of order $k$, so we shall expand the fractions in (3.5,8,11,16,22)
to first order in $R$ and $\Lambda_k$ and evaluate the coefficients
with heat kernel methods.

Let us begin by examining the contribution of the scalar fields.
In the case $m_S\ll k$ in the r.h.s. of (3.4) we can expand the denominator
to first order in $m_S^2$;
there remains to evaluate $\Tr{\partial_t\pk\over\pk}$
and $\Tr{\partial_t\pk\over\pk^2}$. The trace of a function
of $-\lap$ can be expanded in powers of the curvature, with coefficients given
by the Mellin transform of the function itself (see Appendix A).
The result is
$$
\partial_t\Gamma_k|_{\rm scalar}=
{n_S\over2}{1\over(4\pi)^2} \int d^4x\sqrt g
\left[Q_2\left(\partial_t\pk\over\pk\right)
-m_S^2 Q_2\left(\partial_t\pk\over\pk^2\right)+
{1\over6} R Q_1\left(\partial_t\pk\over\pk\right)
\right]\ ,
\eqno (4.1)
$$
where for any function $f$ of the operator $-\lap$, the integral $Q_n(f)$
is defined by 
$$
Q_n(f)={1\over\Gamma(n)}\int_0^\infty dz z^{n-1}f(z)\ .
\eqno(4.2)
$$
The integrals appearing in (4.2) can be evaluated explicitly and are given in
Appendix B.

The contribution of the other fields can be computed in a similar manner.
Whenever there are terms of order $R$ in the denominator one has first
to expand the fraction to first order in $R$, and then evaluate the remaining
traces using formula (A.10). We have:
$$
\eqalign{
\partial_t\Gamma_k|_{\rm matter}=
{1\over(4\pi)^2}\int d^4x\sqrt g
\Biggl\{&
\left({1\over2}n_S-2n_D+n_M+{3\over2}n_P-4n_{RS}\right)
Q_2\left(\partial_t\pk\over\pk\right)\cr
&
+\left(-{1\over2}n_S m_S^2+2n_D m_D^2-{3\over2}n_P m_P^2+4n_{RS}m_{RS}^2\right)
Q_2\left(\partial_t\pk\over\pk^2\right)\cr
+& 
R\Biggl[\left({1\over12}n_S-{1\over3}n_D+{1\over24}n_M+{1\over8}n_P\right)
Q_1\left(\partial_t\pk\over\pk\right)\cr
&
+\left({1\over2}n_D-{3\over8}n_M-{3\over8}n_P+{4\over3}n_{RS}\right)
Q_2\left(\partial_t\pk\over\pk^2\right)\Biggr]
\Biggr\}\cr}\ .
\eqno (4.3)
$$

In a similar way one can calculate the gravitational contribution,
which is
$$
\eqalign{
\partial_t\Gamma_k|_{\rm gravity}={1\over(4\pi)^2}\int d^4x \sqrt{g}
\Biggl\{&
Q_2\left(\partial_t \pk\over \pk\right)
+(6+4\alpha)\Lambda_k Q_2\left(\partial_t \pk\over \pk^2\right)\cr
&
-{13\over24}RQ_1\left(\partial_t \pk\over \pk\right)
-\left({55\over24}+\alpha\right) R Q_2\left(\partial_t \pk\over
\pk^2\right) \cr
&+{\partial_t Z_k\over Z_k}\Bigl[
5 Q_2\left(\pk+\lap\over \pk\right)
+(6+4\alpha)\Lambda_k Q_2\left(\pk+\lap\over \pk^2\right)\cr
&
-{1\over8}RQ_1\left(\pk+\lap\over \pk\right)
-\left({25\over24}+\alpha\right) R Q_2\left(\pk+\lap\over \pk^2\right)
\Bigr]\Biggr\}\cr}\ ,
\eqno(4.4)
$$
where we have neglected terms of order $R\Lambda_k$.

Comparing (4.3) and (4.4) with the formula
$$
\partial_t\Gamma_k=\int d^4x\sqrt{g}
\left[2\partial_t(\kappa_k\Lambda_k)-\partial_t\kappa_k R\right]
\eqno(4.5)
$$
we can read off the complete off shell beta functions for Newton's constant
and for the cosmological constant. For Newton's constant
$$
\partial_t\kappa_k= {a_1 k^2 + \eta_k a_2 k^2}\ ,
\eqno(4.6)
$$
where $\eta_k={\partial_t\ln Z_k}$ can be thought of as the anomalous
dimension of the graviton and
$$
\eqalign{
a_1=&{1\over(4\pi)^2}\left[
\left(-{1\over12}n_S+{1\over3}n_D-{1\over24}n_M-{1\over8}n_P
+{13\over24}\right) q_1^1
\!+\!\left(-{1\over2}n_D+{3\over8}n_M+{3\over8}n_P-{4\over3}n_{RS}
+{55\over24}+\alpha\right)
q_2^2\right];
\cr
a_2=&{1\over(4\pi)^2}\left[
{1\over8} \tilde q_1^1\!+\!\left({25\over24}+\alpha\right)
\!\tilde q_2^2\right];
\cr}\eqno (4.7)
$$
Similarly for the cosmological constant
$$
\partial_t\Lambda_k={1\over \kappa_k}\left(c_1k^4 + c_2 \Lambda_kk^2 +
\sum_i d_i m_i^2k^2
+\eta_k(c_3k^4+c_4\Lambda_kk^2)\right)\ ,
\eqno(4.8)
$$ 
where $i=S,D,P,RS$ and
$$
\eqalign{
c_1=&{1\over(4\pi)^2}\left[
\left({1\over4}n_S-n_D+{1\over2}n_M+{3\over4}n_P-2n_{RS}+{1\over2}\right)
q_2^1\right]\ ;
\cr
c_2=&{1\over(4\pi)^2}\left[
\left({1\over12}n_S-{1\over3}n_D+{1\over24}n_M+{1\over8}n_P-{13\over24}\right)
q_1^1
+\left({1\over2}n_D-{3\over8}n_M-{3\over8}n_P
+{4\over3}n_{RS}+{17\over24}+\alpha\right) q_2^2\right];
\cr
c_3=&{1\over(4\pi)^2}{5\over2}\tilde q_2^1\ ;
\cr
c_4=&{1\over(4\pi)^2}\left[
-{1\over8}\tilde q_1^1
+\left({47\over24}+\alpha\right)\tilde q_2^2\right]\;
\cr
d_S=&-{1\over(4\pi)^2}{1\over4}n_Sq_2^2\ ;\qquad
d_D={1\over(4\pi)^2}n_Dq^2_2\ ;\qquad
d_P=-{1\over(4\pi)^2}{3\over4}n_Pq^2_2\ ;\qquad
d_{RS}={1\over(4\pi)^2}2n_{RS}q^2_2
\cr}
\eqno(4.9)
$$

Let us find the simplest solution to these equations.
Since we are assuming $k^2\gg |R|$, with $R$ finite, 
we cannot use these off-shell equations for arbitrarily
small $k$. So we will have to integrate them starting from
some finite value $\bar k$, where our variables have initial values
$\bar\kappa$ and $\bar\Lambda$.

In addition the effective theory we are using here is only valid
much below Planck's mass, so we also assume
$\bar\kappa\gg k^2$. We get as a solution
$$
\kappa_k \approx {\bar \kappa}+{1\over2}a_1(k^2 -\bar k^2)\ .
\eqno(4.10)
$$
Within this approximation $\eta_k=a_1 {k^2\over\bar\kappa}\ll 1$.
Thus $\kappa_k$ does not change much as $k$ runs over its allowed
range. Since we are assuming that we are not too far off shell, 
$|\Lambda_k|\approx |R|\ll k^2$, so in (4.8)
we can neglect the terms with coefficients
$c_2$ and $c_4$ with respect to those with coefficients $c_1$ and $c_3$.
We then have
$$
\Lambda_k
\approx
\bar\Lambda+{1\over4}{c_1\over\bar\kappa}(k^4-\bar k^4) 
\eqno (4.11)
$$
Remarkably, even though $\Lambda_k$ has dimension of mass squared,
it scales like $k^4$. This is a consequence of the fact that the coefficient
of $R^0$ in the Lagrangian is $\kappa_k\Lambda_k$ and $\kappa_k$ is almost
constant.
The consistency of the approximation requires that $\bar\Lambda\ll k^2$.

For a scalar field, the coefficient $c_1$ that determines the running of 
the cosmological constant is positive. This sign can be understood in terms
of the Casimir effect: raising the IR cutoff removes normal modes and
since the vacuum energy of bosonic oscillators is positive, it lowers
the vacuum energy. For a rigourous argument in flat space in the
context of the Wilsonian RG see [13].
The effect has opposite sign for fermions.
The relative weight of the various fields in $c_1$ is just their number
of degrees of freedom (counted by $b_0$). 
Thus $c_1$ is proportional to the difference of
the total numbers of bosonic and fermionic degrees of freedom.
This may turn out to be positive or negative, depending on the theory.

If $c_1>0$, if we take $\bar k=0$ and if we assume $\bar\Lambda=0$, 
we have an RG trajectory
that lies entirely within the domain of validity of the approximation.
On this trajectory $\Lambda_k$ changes by a factor $10^{240}$ as $k$
varies between the inverse Planck time and the present value of the
Hubble constant. Neighboring trajectories will have similar behaviour for
sufficiently large $k$. We shall discuss these results further in section 7.

\goodbreak
\medskip
\leftline{\bf 5. The large curvature (low energy) limit}
\smallskip
In this section we assume $R>0$. Because Euclidean de-Sitter space is a four
dimensional sphere, in the limit $k^2\ll R$ the scale probed is
larger than the universe itself.  The spectrum of the operators involved is
discrete, and even the lowest eigenvalue is of order $R$ (see Table A.I).  
When the IR cutoff is much smaller than $R$, 
varying it does not change anything,
since all nonzero modes have already been taken into account.  So the
only possible effects must come from the zero modes. 
Since the functions $f(z)$ appearing in the traces in (3.10)
are exponentially decaying functions of $z$ when $z>k^2$, in the limit
$k^2/R\to 0$ the traces are equal to $f(0)$ if the operator has a zero
mode, and zero if the operator does not have zero modes.  The operator
$-\lap$ on the sphere has zero eigenvalues only when acting on
scalars, so the extreme low energy behaviour will be dominated by the
scalar sector: a priori, we could have contributions from
the scalar fields $\phi$, the longitudinal part of the
photon, the electromagnetic ghost, the components $\sigma$ and $h$ 
of the metric fluctuation and the longitudinal part of the gravitational ghost.
However, as noted in section 3, the zero mode should be removed from the trace
over the longitudinal photon, the electromagnetic ghost 
and the longitudinal gravitational ghost.
Thus we could only have contributions from genuine massless scalars and from
the trace of the graviton.

For $z\to 0$ we have:
$$
\pk(z)\to {k^2\over 2a}\left({z\over k^2}\right)^{1-b}
\left(1+O\left({z\over k^2}\right)^b\right)\ ;\qquad\qquad
\partial_t\pk(z)\to {bk^2\over a}\left({z\over k^2}\right)^{1-b}
\left(1+O\left({z\over k^2}\right)^b\right)
\eqno(5.1)
$$
When $b>1$ these functions are divergent at
$z=0$.  In this case the dominant terms in the scalar traces will be
the ones containing the highest powers of $\pk$. We thus find
$$
\partial_t\Gamma_k\to b(\tilde n_S+1)+O(\eta)\ ,\eqno(5.2)
$$
where $\tilde n_S$ is the number of massless scalars.
\footnote{$^*$}
{The result is not continuous in the limit $b\to 1$.
For a gaussian cutoff ($b=1$, $a=1$), $\pk(0)=k^2/2$ and
$\partial_t\pk(0)=k^2$ are finite and we can neglect them with
respect to $R$ and $\Lambda_k$. All terms in (3.22) are of order $k^2/R$
or $k^2/\Lambda_k$, and therefore are negligible.}
The solution 
$\Gamma_k=\bar\Gamma+b(\tilde n_S+1)\ln\left({k\over\bar k}\right)$
has logarithmic infrared divergences, whose meaning we will discuss in the 
next section.

The most difficult regime to study is the one where $k^2\approx\Lambda_k$.
If we start from $k=0$ and let $k^2$ grow, we have to use the
expansion (A.12) and take into account progressively higher modes.
The first correction to (5.2) will come from the first nonzero modes
of $-\lap$.
Using (3.5,8,11,16,22) and Table A.I and neglecting $\eta$, it has the form:
$$
\eqalign{
\partial_t\Gamma_k=&
{35\over2}{\partial_t \pk({2\over3}R)\over \pk({2\over3}R)
+{2\over3}R-2\Lambda_k}
+{10\over2}{\partial_t \pk({1\over4}R)\over\pk({1\over4}R)
+{2\alpha-1\over4}R-2\alpha\Lambda_k} 
+{5\over2}{\partial_t \pk({1\over3}R)\over 
\pk({1\over3}R)-{4\alpha\over{3\alpha-1}}\Lambda_k} \cr &
-10{\partial_t\pk({1\over4}R)\over\pk({1\over4}R)-{1\over4}R}
-5{\partial_t\pk({1\over3}R)\over\pk({1\over3}R)-{1\over2}R} 
-{40\over2}n_{RS}{\partial_t \pk({3\over4}R)\over 
\pk({3\over4}R)+{1\over3}R+m_{RS}^2}
\cr &
+{1\over2}n_M\left[
10{\partial_t \pk({1\over4}R)\over \pk({1\over4}R)+{1\over4}R}
-5{\partial_t\pk({1\over3}R)\over\pk({1\over3}R)}\right]
+{10\over2}n_P{\partial_t \pk({1\over4}R)\over\pk({1\over4}R)+{1\over4}R+m_P^2}
\cr &
-{8\over2}n_{D}{\partial_t\pk({1\over3}R)\over\pk({1\over3}R)
+{1\over4}R+m_D^2}
+{5\over2}n_S{\partial_t\pk({1\over3}R)\over\pk({1\over3}R)+m_S^2}\ .
\cr}
\eqno(5.3)
$$
Each term in this sum is a function of two variables, 
$\Lambda_k/k^2$ and $R/k^2$. When $R/k^2$ becomes large
each term tends exponentially to zero. Since the $n$-th eigenvalue
is of order $n^2 R$, the contributions of higher modes are very
effectively suppressed. We shall examine these contributions
further in the next section.

\medskip
\leftline{\bf 6. Going on shell}
\smallskip
\noindent
One worrying aspect of our calculations is the 
dependence of the results on the gauge parameter $\alpha$.
One notes that even the leading coefficient $a_1$, which appears in the
beta function for $\kappa$ in the high energy regime, is $\alpha$-dependent.
Similarly, in the low energy regime, the contribution of the first nonzero 
mode (5.3) is $\alpha$-dependent.
This phenomenon is common to all calculations of off-shell effective
actions, and is due to the fact that we do not compute physical
quantities. Physical observables would turn out to be
$\alpha$-independent.

There are several possible attitudes in this respect.
One is to note that in practice the value of $a_1$ hardly matters.
If one compares the value of Newton's constant at a planetary scale
(or for that matter at a cosmological scale)
with its value at a scale of a few centimeters,
as relevant for example in a Cavendish experiment, one finds 
$\Delta G/G\approx 10^{-66}a_1$, which is utterly negligible anyway.
This observation however only avoids the question.
A second, more satisfactory answer would be to reformulate the theory 
in a way that is gauge independent.
For example one could compute the running of the Vilkoviski-de Witt
effective action, which is gauge independent by construction.
In de Sitter space, the Vilkoviski-de Witt effective action is obtained
by simply putting $\alpha=0$ [4].

A third attitude, that we will assume in what follows, 
is to try and compute the beta function of other, physically
more meaningful quantities. To guess what the right quantities may be, 
we recall the well-known fact that in one loop calculations
the on shell effective action is gauge independent.
In the present context, going on shell means just putting $R=4\Lambda$.
Since the exact equation that we are using has the form of an improved
one-loop equation, we anticipate that the same may happen here.
If we consider the general truncated RG equation,
the matter contribution is $\alpha$-independent anyway and 
we see from (3.22) that in the leading terms of the gravitational contribution
$\alpha$ cancels out on shell,
while in the term involving $\eta$ this does not generally happen.
However in the high energy regime, if we put $R=4\Lambda_k$ in (4.4) we find
that $\alpha$ cancels:
$$
\eqalign{
\partial_t\Gamma_k|_{\rm gravity, on shell}=&
{24\pi^2\over\Lambda_k^2}
\Biggl\{
2 Q_2\left(\partial_t \pk\over \pk\right)
-{13\over3}\Lambda_kQ_1\left(\partial_t \pk\over \pk\right)
+{19\over3}\Lambda_k Q_2\left(\partial_t \pk\over \pk^2\right) \cr
&+\eta_k\Bigl[ 5Q_2\left(\pk+\lap\over \pk\right) -{1\over2}\Lambda_k
Q_1\left(\pk+\lap\over \pk\right) +{11\over6}\Lambda_k
Q_2\left(\pk+\lap\over \pk^2\right)
\Bigr]\Biggr\}\cr}\ .
\eqno(6.1)
$$

In the low-energy regime, the zero mode term was $\alpha$ independent
and in (5.3), the contribution of the first nonzero mode becomes 
$\alpha$-independent on shell.

What is the physical significance of these results?
The RG equation can be thought of as a vectorfield in the space of all 
coupling constants of the theory, defining the change of these 
couplings under a variation of some external parameter (here $k$). 
Suppose we want to calculate the value of this vector at a point in 
the space of couplings. If we use the background field method, 
we should use a background that solves the (quantum-corrected)
field equations {\it at that point in the space of couplings}.
The field equations will generally depend on the coupling constants, and
therefore one should use a different background at each point in parameter
space. We call the resulting RG equation the {\it on shell RG equation}.
In the present case the effective action is postulated to be of the form
(3.2-3) and the equation of motion says that $R=4\Lambda_k$
(Note that $\kappa_k$ does not appear in this equation.
We can assume that $\kappa$ is actually a constant).
If we now fix a value for $\Lambda_k$, the field equation determines $R$;
this in turn gives the spectrum of the Laplacian and via the equations
given before, the variation of $\Lambda_k$.
The solutions of the on shell RG equation are flow lines, along which
the background field evolves together with the coupling constants.
This on shell RG was applied before in a cosmological context [14].

The meaning of the on-shell RG equation can be further
illuminated by the following observation.
Using that the volume of the four-sphere with curvature scalar $R$ is 
$\int d^4x\sqrt{g}=24{16\pi^2\over R^2}$, we get on shell
$$
\partial_t\Gamma_k|_{\rm on shell}=
\int d^4x\sqrt{g}\left[2\partial_t\Lambda_k\kappa_k
-2\partial_t\kappa_k\Lambda_k\right]
=-48\pi^2\partial_t\left({\kappa_k\over\Lambda_k}\right)
\eqno(6.2)
$$
Therefore, the combination of the coupling constants whose beta function is
$\alpha$-independent is their dimensionless ratio $F_k=\kappa_k/\Lambda_k$.
This fits with the observation that in pure gravity
$\Lambda$ and $\kappa$ by themselves are not measurable, 
but their ratio is [15]. 

Let us now review the solutions of the on-shell RG equation.
If we compute the beta function of $F_k$ in the high energy regime
we get from (4.6-9): 
$$
\partial_t F_k ={F_k^2\over\kappa_k^2}\left(-c_1k^4+(a_1-c_2)\Lambda_kk^2-
\sum_i d_i m_i^2k^2 +\eta_k(-c_3k^4+(a_2-c_4)\Lambda_kk^2)\right)\ .
\eqno(6.3) 
$$ 
One observes again that $\alpha$ cancels out in the combinations
$a_1-c_2$ and $a_2-c_4$. This equation is still $\alpha$-dependent, but
only in the subleading order. Within the approximation of eqs.(4.10,11) we
have 
$$ 
\partial_t F_k=-c_1{k^4\over\kappa_k^2}F_k^2
\eqno(6.4)
$$ 
This gives back (4.11), which is therefore a trustworthy result. 

In the low energy regime, we have from (5.2) and (6.2)
$$
\partial_t F_k=-{(1+\tilde n_S)b\over 48\pi^2} \eqno(6.5)
$$
which yields
$$
\Lambda_k={\bar\Lambda\over 
1-{(2+\tilde n_S)b\over 48\pi^2}{\bar\Lambda\over\bar\kappa}
\ln{k\over\bar\kappa}} \eqno(6.6)
$$
For $k\to 0$ we then have $\Lambda_k\to 0$. 
This running is very slow, however.

Let us now examine the contribution of the first nonzero modes.
Neglecting the contribution of massive particles, (5.3) becomes on 
shell: 
$$
\eqalign{
\partial_t\Gamma_k|_{\rm on shell}=&
{35\over2}{\partial_t \pk({8\over3}\Lambda_k)\over \pk({8\over3}\Lambda_k)
+{2\over3}\Lambda_k}
-{10\over2}{\partial_t \pk(\Lambda_k)\over\pk(\Lambda_k)-\lambda_k}
+{10\over2}n_M
{\partial_t \pk(\Lambda_k)\over \pk(\Lambda_k)+\Lambda_k}
-{5\over2}n_M{\partial_t\pk({4\over3}\Lambda_k)\over\pk({4\over3}\Lambda_k)}
\cr &
-{8\over4}n_{W}{\partial_t\pk({4\over3}\Lambda_k)\over\pk({4\over3}\Lambda_k)
+\Lambda_k}
+{5\over2}\tilde n_S{\partial_t\pk({4\over3}\Lambda_k)\over\pk({4\over3}\Lambda_k)} \cr}
\eqno(6.7)
$$
Each fraction in this sum goes exponentially to zero for $\Lambda_k/k^2>1$
and is equal to $b$ for $\Lambda_k/k^2=0$. The shape of the function 
is similar to a damped exponential for $b=1$ and tends to a step function
for large $b$. The $i$-th modes will give similar contributions, with the
argument in $\pk$ and $\partial_t\pk$ of order $i^2\Lambda_k$ and
the overall coefficient of order $i^3$.
One understands from here the qualitative behaviour of the solutions.
Putting the cutoff at momentum $k$ means summing the first $i$ modes,
with $i^2=k^2/\Lambda_k$.
Adding higher modes will pile up more of these functions, so for
$\Lambda_k/k^2<1$ the r.h.s. will grow fast with $k$, roughly like
$i^4$, while
for $\Lambda_k/k^2>1$ only the sero mode contribution remains, and it
is independent of $\Lambda_k/k^2$.

\medskip
\leftline{\bf 7. Discussion}
\smallskip
\noindent
In this paper we have computed the scale dependence of the gravitational 
action as a function of an external parameter $k$, having the physical
significance of resolution of the apparatus that is used to make the
measurements. Features of the system with length scales smaller than
$k^{-1}$ cannot be observed and are therefore integrated over.
Their effect is summarized in the effective action $\Gamma_k$, that
describes the dynamics of the modes with wavelengths larger than $k^{-1}$.

In a gravitational context, the renormalization group is often seen
as the behaviour of renormalized quantities under global (constant)
rescalings of the metric [16]. We will now relate this point of view to
the one used in this paper.
Let us consider for the sake of generality an action functional
$S(g_{\mu\nu},\phi,\lambda)$, where $g_{\mu\nu}$ is a background
metric, $\phi$ are quantum fields and $\lambda$ are coupling
constants. Under a global scaling of the metric $g_{\mu\nu}\to\omega^2
g_{\mu\nu}$, the action is invariant provided all fields and coupling
constants are scaled according to their canonical mass dimension
$$
S(\omega^2 g_{\mu\nu},\phi,\lambda)= S(g_{\mu\nu},\omega^\dphi
\phi,\omega^\dlambda \lambda)
\eqno(7.1)
$$
In fact this relation can be taken as the definition of the canonical
dimensions $\dphi$ and $\dlambda$.

By following the definition of the average effective action given in
the preceding section, one can easily prove the following equality:
$$
\Gamma_k(\omega^2g_{\mu\nu},\phi,\lambda)
=\Gamma_{\omega k}(g_{\mu\nu},
\omega^\dphi\phi,\omega^\dlambda\lambda)
\eqno(7.2)
$$
Note that $k$ behaves like any coupling constant with dimension of
mass.  This formula, which follows simply from dimensional analysis,
relates the behaviour of the effective action under changes of $k$ to
the behaviour under scalings of the metric.

For $\omega\to \infty$ the distance between two given
points in spacetime goes to infinity, so this limit is called the infrared
limit and vice versa the limit $\omega\to 0$ is called the ultraviolet
limit. On the other hand
the Wilsonian renormalization group describes the behaviour of the
theory as $k$ is varied and one calls the limits
$k\to 0$ and $k\to\infty$ the infrared and ultraviolet limit respectively.
One sees from eq. (6.2) that for example $\omega\to\infty$ is
equivalent (modulo rescalings of the fields and coupling constants) to
the limit $k\to\infty$ and thus the terminology of what constitutes
the ultraviolet and the infrared is reversed with respect to
considerations of scaling the metric.  (This was implicitly stated in
[14], eq.(23c) where a scalar field theory was shown to have an
ultraviolet fixed point for $\xi=1/6$, while in the literature [16]
this is called an infrared fixed point)

We have already discussed the dependence of the results on
the gauge parameter $\alpha$. There is an additional
dependence of the results on arbitrary parameters, namely
on the parameters $a$
and $b$ that define the shape of the cutoff function (2.4).
In this respect, 
one observes that the numerical value of all integrals (given
in  Appendix B) depends on the parameters $a$ and $b$.
The running of the couplings is therefore not universal. 
This is in accordance with the fact that Einstein gravity can be
thought of as a spontaneusly broken $GL(4)$ gauge theory, with
a symmetry breaking scale of the order of Planck's energy [17].
The renormalization group flow of coupling constants in spontaneously
broken theories below the breaking scale is always non-universal.
It is therefore reasonable to expect that gravity will exhibit
universal properties only beyond the Planck scale.

Even though we do not have a completely consistent theory of gravity
in the ultraviolet, one can apply the methods of this paper to other
effective theories of gravity that could be relevant in that regime.
The running of Newton's constant was calculated before in a
gauge theory of gravity with torsion and dilaton, 
containing terms quadratic in the curvature [18]. 
The results obtained here are in agreement with those of [18]
in their domain of applicability. 
One thus gets a picture of a dilatonic theory
in which the v.e.v. of the dilaton runs linearly above Planck's energy
and settles to an almost constant value as the energy drops below
the Planck threshold. This value can be identified with the (almost constant)
Newton constant discussed in the present paper.
We mention that the gravitational contributions to the running of a scalar
self-coupling was computed using similar methods in [19]. 

Another intriguing possibility is that the effective action for
gravity at low energies is non-local as suggested by 
several authors in the past [1,4,20].
In particular in [4] it was assumed that the effective action could contain
a term of the form $V\ln V$, where $V$ is the four-dimensional
volume. The effect of this term would be a highly nonlinear
relation between the bare and effective cosmological constant.
It is possible that our calculation, though restricted to local
Lagrangians, may contain some hints of similar effects.
More precisely, the running cosmological constant discussed here
should be identified, for $k\to 0$, with $\Lambda_{\rm eff}$.
We also saw in Sect.6 that in the IR limit terms like $R^{-1}$ 
must appear. The existence of similar nonlocal terms could have profound
consequences at the cosmological level.
\bigskip
This work is supported in part by GRANT NO. ERBFMRXCT 960090
\bigskip

\leftline{\bf Appendix A: Spectral geometry on the sphere}
\smallskip
\noindent

The spectrum of the operator $-\lap$ on the sphere, acting on fields
of definite spin, can be computed from the eigenvalues of the Casimir
operator of the group $SO(5)$ on irreducible representations [11]. 
Table I collects the results that are used in this paper:
\medskip
\newbox\sstrutbox
\setbox\sstrutbox=\hbox{\vrule height20pt depth15pt width0pt}
\def\bigstrut{\relax\ifmmode\copy\sstrutbox\else\unhcopy\sstrutbox\fi}
\relax
\vbox{\tabskip=0pt \offinterlineskip 
\def\tablerule{\noalign{\hrule}}
\halign to 420pt {\strut#& \vrule#\tabskip=1em plus 2em&
\hfil#\hfil& \vrule#& 
\hfil#\hfil& \vrule#& 
\hfil#\hfil& \vrule#
\tabskip=0pt \cr\tablerule
&&\multispan5\hfil\bigstrut TABLE A.I: Eigenvalues of
$-\nabla^2$ on the sphere \hfil&\cr\tablerule
&& Spin \bigstrut && Eigenvalues && Multiplicity &
\cr\tablerule
&&0\bigstrut&& $\lambda_i^{(0)}={1\over12}Ri(i+3)$\ ;\ $i$=0,1,... &&
${1\over6}(i+1)(i+2)(2i+3)$ &\cr\tablerule
&&1/2\bigstrut&& $\lambda_i^{(1/2)}={1\over12}R(i+1)^2$\ ;
\ $i$=1,... && ${2\over3}i(i+1)(i+2)$
&\cr\tablerule
&&1\bigstrut&& $\lambda_i^{(1)}={1\over12}R(i^2+3i-1)$\ ;\ $i$=1,... &&
${1\over2}i(i+3)(2i+3)$  &\cr\tablerule
&&3/2\bigstrut&& $\lambda_i^{(3/2)}={1\over12}R(i+1)^2$\ ;\ $i$=2,... &&
${4\over3}(i-1)(i+1)(i+3)$  &\cr\tablerule
&&2\bigstrut&& $\lambda_i^{(2)}={1\over12}R(i^2+3i-2)$\ ;\ $i$=2,... &&
${5\over6}(i-1)(i+4)(2i+3)$  &\cr\tablerule
\hfil\cr}}
\relax
\medskip

In this table the spin $1/2$ and spin $3/2$ fields are irreducible
and therefore chiral. For Dirac spinors and nonchiral Rarita-Schwinger fields
the multiplicities have to be doubled.
Formally for a function of $-\lap$ we have 
$\Tr f(-\lap)=\sum_i f(\lambda_i)$ where $\lambda_i$ 
are the eigenvalues of $-\lap$.
The best studied such trace is the trace of the heat kernel $e^{-s(-\lap)}$.
For small $s$ it has an expansion of the form:
$$
Tr e^{-s(-\lap)}=B_0(-\lap) s^{-2}+B_2(-\lap)
s^{-1} +B_4(-\lap)+....
\eqno(A.1)
$$
where $B_n=\int d^4x \sqrt{g}\tr b_n$ and $b_n$ is a polynomial
tensor involving the $n/2$-th power of the curvature. The coefficients
of these polynomials depend on the space the operator is acting on.
For operators acting on unrestricted scalar, vector, tensor and spinor
fields, the coefficients can be computed using the formula [21]:
$$
\eqalignno{
&b_0={1\over(4\pi)^2}{\bf 1} &(A.2a)\cr
&b_2={1\over(4\pi)^2}{R\over6}{\bf 1} &(A.2b)\cr
&b_4={1\over(4\pi)^2}\biggl[
\left({1\over180}R_{\mu\nu\rho\sigma}R^{\mu\nu\rho\sigma}
-{1\over180}R_{\mu\nu}R^{\mu\nu}
+{1\over72}R^2
+{1\over30}\nabla_\mu\nabla^\mu R\right){\bf 1}
+{1\over12}{\cal F}_{\mu\nu}{\cal F}^{\mu\nu}\biggr]\ ,&(A.2c)\cr}
$$
where ${\cal F}_{\mu\nu}$ is defined as the commutator of covariant
derivatives acting on the field in question.

For fields subject to differential constraints, such as the spin 1, 3/2 and 2
fields $A_\mu^T$,
$\Phi_\mu^T$ and $h_{\mu\nu}^T$, the coefficients can be computed from the
coefficients of the differentially unrestricted fields, using
the decompositions (3.12), (3.17) and Table A.I.

We give here a derivation appropriate to the Euclidean four-dimensional
sphere. On a vector field $A_\mu$, using the formula 
$$
-\lap\nabla_\mu\phi=\nabla_\mu\left(-\lap-{1\over4}R\right)\phi
\eqno(A.3)
$$ 
one finds that
$$
Tre^{-s(-\lap)}|_{A_\mu}=
Tre^{-s(-\lap)}|_{A_\mu^T}+
Tre^{-s(-\lap-{1\over4}R)}|_\phi-e^{sR/4}
\eqno(A.4)
$$
The last term (which in the expansion (A.1) would appear as a $-1$
in the coefficient $B_4$) is due to the fact that the operator 
$-\lap-{1\over4}R$ acting on 
scalars has a negative mode $\phi_0$=constant which does not correspond to any 
normal mode of the operator $-\lap$ on vectors.  
From (A.4) one can compute $B_n(-\lap)|_1$.
Similarly for the Rarita-Schwinger field using
$$
-\lap\left(\nabla_\mu-{1\over4}\gamma_\mu\gamma^\nu\nabla_\nu\right)\xi=
\left(\nabla_\mu-{1\over4}\gamma_\mu\gamma^\nu\nabla_\nu\right)
\left(-\lap-{1\over3}R\right)\zeta
\eqno(A.5)
$$
one has
$$
Tre^{-s(-\lap)}|_{\Psi_\mu}=
Tre^{-s(-\lap)}|_{\Phi_\mu^T}+
Tre^{-s(-\lap)}|_{\psi}+
Tre^{-s(-\lap-{1\over3}R)}|_\zeta-8
\eqno(A.6)
$$
The last term corresponds to the eight zero modes of $-\lap-{1\over3}R$
on spinors that do not correspond to normal modes of $-\lap$ on $\Phi_\mu$.
For the graviton we use the formulae:
$$
-\lap\left(\nabla_\mu\xi_\nu+\nabla_\nu\xi_\mu\right)=
\nabla_\mu\left(-\lap-{5\over12}R\right)\xi_\nu+
\nabla_\nu\left(-\lap-{5\over12}R\right)\xi_\mu
\eqno(A.7)
$$ 
and
$$
-\lap\left(\nabla_\mu\nabla_\nu-{1\over4}g_{\mu\nu}\lap\right)\sigma=
\left(\nabla_\mu\nabla_\nu-{1\over4}g_{\mu\nu}\lap\right)
\left(-\lap-{2\over3}R\right)\sigma
\eqno(A.8)
$$
to obtain
$$
\eqalign{
Tre^{-s(-\lap)}|_{h_{\mu\nu}}=&
Tre^{-s(-\lap)}|_{h_{\mu\nu}^T}+
Tre^{-s(-\lap-{5\over12}R)}|_\xi+
Tre^{-s(-\lap)}|_h+
Tre^{-s(-\lap-{2\over3}R)}|_\sigma\cr
&\qquad\qquad\qquad -e^{2sR/3}-5e^{sR/3}-10e^{sR/6}\cr}
\eqno(A.9)
$$
The last term
comes from the ten negative modes of the operator
$-\lap-{5\over12}R$ on $\xi$, which are the Killing vectors of $SO(5)$
and therefore do not correspond to normal modes of $h_{\mu\nu}$
($\nabla_\mu\xi_\nu+\nabla_\nu\xi_\mu=0$); the second last term
comes from the five negative
modes of $-\lap$ (with eigenvalue $-{1\over3}R$)
proportional to the functions $x^a$, the coordinates of
a flat space in which the sphere is embedded. These functions
satisfy $4\nabla_\mu\nabla_\nu x^a=g_{\mu\nu}\lap x^a$;
the third last term comes from
the (constant) negative mode of $-\lap-{2\over3}R$ on scalars,
(with eigenvalue $-{2\over3}R$).
\footnote{$^*$}
{The presence of the constant terms in (A.4), (A.6) and (A.9) can be
confirmed by explicit calculations of the coefficient $B_4$ of the
l.h.s., and making a comparison to the zeta functions of the 
operators appearing on the r.h.s., whose spectra are known.}
In this way one can compute the relevant heat kernel coefficients
for $-\lap$ acting on irreducible representations of $SO(5)$:
\medskip
\newbox\sstrutbox
\setbox\sstrutbox=\hbox{\vrule height20pt depth15pt width0pt}
\def\bigstrut{\relax\ifmmode\copy\sstrutbox\else\unhcopy\sstrutbox\fi}
\relax
\vbox{\tabskip=0pt \offinterlineskip 
\def\tablerule{\noalign{\hrule}}
\halign to 420pt {\strut#& \vrule#\tabskip=1em plus 2em&
\hfil#& \vrule#& 
\hfil#\hfil& \vrule#& 
\hfil#\hfil& \vrule#& 
\hfil#\hfil& \vrule#& 
\hfil#\hfil& \vrule#& 
\hfil#\hfil& \vrule#
\tabskip=0pt \cr\tablerule
&&\multispan{11}\hfil\bigstrut TABLE A.II: Heat kernel
coefficients \hfil&\cr\tablerule 
&&\omit\hidewidth Spin \bigstrut\hidewidth&&
\omit\hidewidth 0 \hidewidth&&
\omit\hidewidth 1/2 \hidewidth&&
\omit\hfil 1 \hfil&&
\omit\hfil 3/2 \hfil&&
\omit\hfil 2 \hfil&
\cr\tablerule
&&\bigstrut ${\rm tr}b_0$
&& 1 && 2 && 3 && 8 && 5 &\cr\tablerule
&&\bigstrut ${\rm tr}b_2$
&& ${1\over6}R$ && ${1\over3}R$ && ${1\over4}R$ && $0$ && $-{5\over6}R$
&\cr\tablerule
\hfil\cr}}
\relax
\medskip
For non-chiral (Dirac) spinors and Rarita-Schwinger fields, the
coefficients in the third and fifth columns have to be doubled.

For a function $f(z)$, one finds the expansion
$$
\Tr f(-\lap)=B_0(-\lap)Q_2(f)+B_2(-\lap)Q_1(f)+B_4(-\lap)Q_0(f)+\ldots
\eqno(A.10)
$$
where
$$
Q_n(f)={1\over\Gamma(n)}\int_0^\infty dz z^{n-1}f(z)\ .
\eqno(A.11)
$$

For large $s$ the trace of the keat kernel is dominated by the
lowest eigenvalues:
$$
Tr e^{-s(-\lap)}=e^{-s\lambda_0}+e^{-s\lambda_1}+....
\eqno(A.12)
$$
and more generally for a function that decays fast enough
for large $s$
$$
Tr f(-\lap)=f(\lambda_0)+f(\lambda_1)+....
\eqno(A.13)
$$
\medskip

\leftline{\bf Appendix B: some integrals}
\smallskip
\noindent
We collect here the result of some integrations which appear in
section 4:

$$
\eqalignno{
Q_1\left({\partial_t\pk\over\pk}\right)=\ &k^2\,q_1^1=
\ k^2{2\over(2a)^{1/b}b}\Gamma\left({1\over b}\right)
\zeta\left(1+{1\over b}\right)\ ;&(B.1)\cr
Q_2\left({\partial_t\pk\over\pk}\right)=\ &k^4\,q_2^1=
\ k^4{4\over (2a)^{2/b}b}\Gamma\left({2\over b}\right)
\zeta\left(1+{2\over b}\right)\ ;&(B.2)\cr
Q_2\left({\partial_t\pk\over\pk^2}\right)=\ &k^2\,q_2^2=
\ k^2{2\over(2a)^{1/b} b}\Gamma\left({1\over b}\right)\ ;&(B.3)\cr
Q_1\left({\pk+\lap\over\pk}\right)=\ &k^2\,\tilde q_1^1=
\ k^2{1\over(2a)^{1/b}b}
\Gamma\left({1\over b}\right)\ ;&(B.4)\cr
Q_2\left({\pk+\lap\over\pk}\right)=\ &k^4\tilde q_2^1=
\ k^4{1\over(2a)^{2/b}b}
\Gamma\left({2\over b}\right)\ ;&(B.5)\cr
Q_2\left({\pk+\lap\over\pk^2}\right)=\ &k^2\tilde q_2^2=
\ k^2{1\over(2a)^{1/b}b}
\left(1-{1\over 2^{1/b}}\right)
\Gamma\left({1\over b}\right)\ ;&(B.6)\cr}
$$

\goodbreak
\centerline{\bf References}
\bigskip
\noindent
\item{1.} J. Donoghue, Phys.Rev. {\bf D 50} 3874 (1994); 
{\it ibid.}{\bf D54} 4963 (1996);
Helv.Phys.Acta {\bf 69} 269 (1996). 
\smallskip
\item{2.} S. Weinberg, Rev. Mod. Phys. {\bf 61}, 1 (1989); 
\smallskip 
\item{3.} S. Coleman, Nucl. Phys. {\bf B 310}, 643 (1988);
\smallskip
\item{4.} T.R. Taylor and G. Veneziano, Nucl. Phys. {\bf B 345}, 210 (1991);
\smallskip
\item{5.} N.C. Tsamis and R.P. Woodard, Ann. of Phys. (NY) {\bf 238}, 1
(1995); \smallskip
\item{6.} C. Wetterich, Phys.Lett {\bf B 301}, 90 (1993);
\smallskip
\item{7.} M. Reuter,  hep-th/9605030;
\smallskip
\item{8.} C. Wetterich, hep-th/9501119;
\smallskip
\item{9.} C. Wetterich and M. Reuter, Nucl.Phy {\bf B 417}, 181 (1994);
\smallskip
\item{10.} G.W. Gibbons and M.J. Perry, Nucl. Phys. {\bf B 146}, 90 (1978);
\smallskip
\item{11.} E.S. Fradkin and A.A. Tseytlin, Nucl. Phys. {\bf B 234},
472, (1984); 
\smallskip
\item{12.} B. Allen, Phys. Rev. {\bf D 34} 3670 (1986).
\smallskip
\item{13.} V. Periwal, Mod. Phys. Lett. {\bf A10}, 1543 (1995);
\smallskip
\item{14.} R.Floreanini and R.Percacci, Phys.Lett. {\bf B 356} 205 (1995).
\smallskip
\item{15.} H. Kawai and M. Ninomiya, Nucl.Phys. {\bf B 336}, 115 (1990).
\smallskip
\item{16.} Nelson et al., Phy.Rev {\bf D 25}, 1019 (1982); 
{\it ibid.} {\bf D 29}, 2750 (1984);\hfil\break
I.L. Buchbinder et al. ``Effective Action in Quantum Gravity'';
\smallskip
\item{17.} R. Percacci, Nucl. Phys. {\bf B 353} 271 (1991);
\smallskip
\item{18.} R.Floreanini,R.Percacci, Phys.Rev. {\bf D 52}, 896 (1995);
\smallskip
\item{19.} L.Griguolo,R.Percacci, Pys.Rev {\bf D 52},(1995)  5787.
\smallskip
\item{20.} C. Wetterich, gr-qc/9704052.
\smallskip
\item{21.} B. de Witt, {\it Dynamical theory of groups and fields},
in ``Relativity, Groups and Topology'', ed C.M. de Witt and B. de Witt,
Gordon and Breach, NY (1964).
\smallskip

\vfil
\eject
\bye